\documentclass{article}
\usepackage[dvips]{graphicx}
\usepackage{bm}
\begin{document}
\begin{center}
{\bf \large WEAK AND STRONG LENSING STATISTICS
\footnote{Talk given at the ISSI workshop {\it Matter in the Universe}, 
19-23 March 2001, Bern-CH}}
\vskip3mm
{\bf \large Norbert Straumann}
\vskip2mm
Institut f\"ur Theoretische Physik, Universit\"at Z\"urich, Switzerland
\end{center}
\vskip4mm
\begin{abstract}
After a brief introduction to gravitational lensing theory, a rough overview 
of the types of gravitational lensing statistics that have been performed so 
far will be given. I shall then concentrate on recent results of galaxy-galaxy 
lensing, which indicate that galactic halos extend much further than can be 
probed via rotation of stars and gas.
\end{abstract}

\section{Introduction}
\label{sec:introduction}
Since I am the first at this meeting who talks about gravitational lensing 
(GL), I thought it might be useful if I start by recalling some of the basics 
of GL-theory and standard terminology.
\\
\indent
Space does not allow me to discuss in any detail the types of lensing 
statistics that have been performed so far. After a brief discussion of the 
main ones, I shall concentrate on some recent studies of galaxy-galaxy 
lensing which have led to some interesting -although not definite- results on 
the properties of galactic halos. The existing measurements demonstrate the 
power and potential of this method. The data indicate that halos of typical galaxies continue an isothermal profile to a radius of at least 260 h$_o^{-1}$ 
kpc, but in the foreseeable future the situation should improve considerably.

\section{Basics of Gravitational Lensing Theory}
\label{sec:basics}

Gravitational lensing has the distinguishing feature of being independent of 
the nature and the physical state of the deflecting mass distributions. 
Therefore, it is perfectly suited to study dark matter on all scales. 
\\
\indent
Moreover, the theoretical basis of lensing is very simple. For all practical 
purposes we can use the ray approximation for the description of light 
propagation. In this limit the rays correspond to null geodesics 
in a given gravitational field. For a qualitative understanding it is helpful 
to use the Hamilton-Jacobi description of ray optics. Let me briefly 
recapitulate how this looks in general relativity (see e.g. Straumann 1999). 
\\
\indent
If we insert the following eikonal ansatz for the Maxwell field
$$
F_{\mu \nu} \, = \, {\cal R}e \Bigl( f_{\mu \nu} \, e^{ i S} \Bigr ) , 
$$
with a slowly varying amplitude $f_{\mu \nu}$ and a real $S$ into the general 
relativistic Maxwell equations, we obtain the eikonal equation
\begin{equation}
g^{\mu \nu} \, \partial_{\mu} S \partial _{\nu} S \, = \, 0 .
\end{equation}
This says that the vector field $k^\mu (x) = \nabla ^\mu S$ is null. The 
integral curves of $k^\mu (x)$ are the light rays. From the general 
relativistic eikonal equation (1) one easily shows that they are -as expected- 
null geodesics. By construction they are orthogonal to the wave fronts 
$S=const$.
\\
\indent
For an almost Newtonian situation the metric is ($c=1$):
\begin{equation}
g_{\mu \nu} dx^ \mu dx^ \nu \, = \, 
-(1 + 2 U) dt^2 + (1-2 U) d \mathbf{x}^2 , 
\end{equation}
where $U$ is the Newtonian potential. Since this is time independent, we can make the ansatz
\begin{equation}
S(x) \, = \, \widehat{S} (\mathbf{x}) - \omega t
\end{equation}
and obtain for $\widehat{S} (\mathbf{x})$ the standard eikonal equation of 
geometrical optics
\begin{equation}
( \bm{\nabla} \widehat{S} )^2 \, = \, n^2 \omega^2 ,
\end{equation}
with the \underline{effective refraction index}
\begin{equation}
n (\mathbf{x}) \, = \, 1 - 2 U (\mathbf{x} ) / c^2 .
\end{equation}
This shows that in the almost Newtonian approximation general relativity 
implies that gravitational lensing theory is just usual ray optics with the 
effective refraction index (5). (In a truely cosmogical context, things are 
not quite that simple).
\\
At this point the qualitative picture sketched in Fig.1 is useful for a first 
orientation. It shows the typical structure of wave fronts in the presence of 
a cluster perturbation, as well as the rays orthogonal to them.

\begin{figure}
\centerline{\includegraphics[clip=,angle=0,width=0.67\textwidth]
{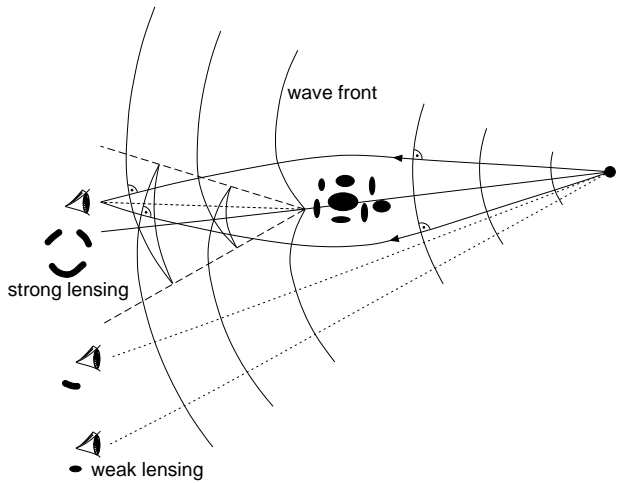}}
\caption{Wave fronts and light rays in the presence of a cluster 
perturbation.}
\end{figure}

For sufficiently strong lenses the wave fronts develop edges and 
self-interactions. An observer behind such folded fronts obviously sees more 
than one image. From this figure one also understands how the time delay of 
pairs of images arises: this is just the time elapsed between crossings of 
different sheets of the same wave front. Since this time difference is, as 
any other cosmological scale, depending on the Hubble parameter, GL 
provides potentially a very interesting tool to measure the Hubble 
parameter. So far, the uncertainties of this method are still quite large, 
but the numbers fall into a reasonable range.
\\
\indent
If the extension of a lens is much smaller than the distances between the lens and the observer as well as the source, one can derive the following 
equation for the lens map (Schneider et al 1992, Straumann 1999): 
Adopting the notation in Fig.2, the map $\varphi$: 
$\bm{\theta} \mapsto \bm{\beta}$
 (image $\rightarrow$ source) is the gradient map:
\begin{equation}
\varphi (\bm{\theta}) \, = \, \bm{\nabla} \Bigl(
\frac{1}{2} \bm{\theta}^2 - \psi(\bm{\theta}) \Bigr) ,
\end{equation}
where the \underline{deflection potential} $\psi$ satisfies the 
two-dimensional Poisson equation
\begin{equation}
\Delta \psi \, = \, 2 \kappa .
\end{equation}
Here, $\kappa$ is the projected mass density $\Sigma (\bm{\theta})$ 
in units of the critical surface mass density. (The latter is defined such that if $\kappa > 1$ somewhere there are always multiple images for some source 
positions.)

\begin{figure}
\begin{center}
\centerline{\includegraphics[clip=,angle=0,width=0.30\textwidth]
{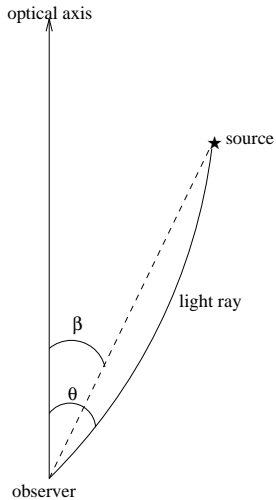}}
\end{center}
\caption{Notation adopted for the description of the lens geometry.}
\end{figure}

The linearized lens map
\begin{equation}
{\cal A} \, = \, \Bigl( \delta_{ij} - \frac{\partial ^2 \psi}{
\partial \theta_i \partial \theta_j} \Bigr) 
\end{equation}
is usually parametrized as  
\begin{equation} 
{\cal A} \, = \, 
\left( \begin{array}{cc}
1- \kappa -\gamma_1 & -\gamma_2 \\
-\gamma_2 & 1-\kappa+\gamma_1
\end{array} \right) .
\end{equation}
Note that the \underline{complex shear} $\gamma = \gamma_1 + i \gamma_2$ 
describes the trace-free part of ${\cal A}$ and does, therefore, \underline{not} transform like a vector.
\\
\indent
Let me also recall how $\gamma$ can be measured in the limit of weak lensing. Knowing the surface brightness distribution $I(\bm{\theta})$ of a 
galaxy image allows us to compute the tensor $Q_{ij}$ of brightness moments 
and thus the \underline{complex ellipticity} 
\begin{equation}
\epsilon \, = \, \frac{Q_{11} - Q_{22} + 2i Q_{12}}
{{\rm tr}Q + 2 {\rm det}Q} .
\end{equation}
By making use of (9) one find for weak lensing ($\kappa \ll 1$) the 
following relation between $\epsilon$ and the corresponding source 
ellipticity $\epsilon^{(s)}: \epsilon = \epsilon^{(s)} + \gamma$. For an 
individual galaxy this is, of course, not of much use, but for a sufficiently 
dense ensemble it is reasonable to assume that the average value 
$<\epsilon^{(s)}>$ vanishes. One can, however, not exclude some intrinsic 
alignments of galaxies caused, for instance, by 
tidal fields. This can be tested, as I shall discuss later. If we set 
$<\epsilon^{(s)} > =0$ we have $\gamma = < \epsilon >$, and $\gamma$ can thus be measured.
\section{Types of Lensing Statistics}
\label{sec:lensing}
I confine myself just to a few remarks on statistics involving 
\underline{strong} lensing, because in all cases the theoretical and 
observational uncertainties are still quite large.
\vskip2mm
(i) In several recent studies (e.g. Kochanek 1996; Chiba and Yoshii 1999, 
Chiba and Futamase 1999; Cheng and Krauss 2000) the statistics of strong 
gravitational lensing of distant quasars by galaxies has recently been 
re-analyzed. Observationally, there are only a few strongly lensed quasars 
among hundreds of objects. The resulting bounds on $\Omega_M$ and 
$\Omega_\Lambda$ are, however, not very tight because of systematic 
uncertainties in the galaxy luminosity functions, dark matter velocity 
dispersions, galaxy core radii and limitations of the observational 
material.
\vskip2mm
(ii) On the basis of existing surveys, the statistics of strongly lensed 
radio sources has been studied in several recent papers (e.g. Falco et al. 
1998; Cooray 1999; Helbig et al. 1999). Beside some advantages for 
constraining the cosmological model, there is the problem that the redshift distribution of the radio sources is largely unknown. (One can, however, make use 
of s strong correlation between the redshift and flux density distributions.)
\vskip2mm
(iii) Clusters with redshifts in the interval $0.2 < z_c < 0.4$ are efficient lenses for background sources at $z_s \sim 1$. For several reasons one can expect that the probability for the formation of pronounced arcs is a sensitive 
function of $\Omega_M$ and $\Omega_\Lambda$. First, it is well-known that 
clusters form earlier in low density universes. Secondly, the proper volume per unit redshift is larger for low density universes and depends strongly on 
$\Omega_\Lambda$ for large redshifts. An extensive numerical study of arc 
statistics has been performed by Bartelmann et al. (1998), with the result 
that the optical depth depends strongly on $\Lambda$. In the semi-analytical 
treatment of Kaufmann and Straumann (2000) only a weak $\Lambda$-dependence 
was, however, found. We compared our theoretical expectations with the results 
of a CCD imaging survey of gravitational lensing from the Einstein Observatory 
Extended Medium-Sensitivity Survey (EMSS). We believe that at least the shape 
of the maximum likehood regions is correct. The absolute numbers are quite 
uncertain at the present stage. Among several empirical parameters $\sigma_8$ 
affects the prediction most strongly. However, a low-density universe is 
clearly favored. 
\vskip2mm
Improvements are possible, but the method will presumably never become precise.
\\
\indent
\underline{Weak lensing} is more promissing because linear perturbation theory is 
sufficiently accurate. The theoretical tools for analysing weak-lensing data 
are well described in a recent review article of Bartelmann and Schneider 
(2001). Since Y. Mellier will talk at this meeting about the gravitational 
lensing caused by large scale structures (which has recently been detected 
by several groups), I shall concentrate below on galaxy-galaxy lensing. 
Before doing this, I should, however, discuss a crucial issue which is 
relevant for both types of weak lensing statistics.

\section{Discrimination of Weak Lensing From Intrinsic Spin Correlations (etc)}
\label{sec:spin}
\indent
The shear field $\gamma$ can only be determined from the observed ellipticity 
$\epsilon$ if 
$<\epsilon^{(s)}>=0$. As already mentionned, this is not 
guaranteed. Therefore it is important to have tests for this statistical 
assumption.
\\
\indent
For an elegant derivation of such a test we consider an arbitrary ellipticity 
field $\epsilon(\bm{\theta})$ as a complex function of $z=\theta_1 
+ i \theta_2$ and decompose it into its {\it electric} and {\it magnetic} 
parts: If $\partial$ and $\overline{\partial}$ denote the Wirtinger 
derivatives 
\\
$\Bigl[ \partial = \frac{1}{2} (\partial_{\theta_1} - i 
\partial_{\theta_2} ), 
\overline{\partial} = \frac{1}{2} (\partial_{\theta_1} + i 
\partial_{\theta_2} ) \Bigr]$, then we can represent $\epsilon$ as 
\begin{equation}
\epsilon \, = \, 2 \overline{\partial}^2 \phi
\end{equation}
by a potential $\phi$. (This is an immediate application of Dolbaut's lemma; 
see Straumann 1997.) 
Decomposing $\phi$ into its real and imaginary parts, $\phi=\phi_E+i\phi_B$, 
provides the electric and magnetic parts of $\epsilon$. If $\epsilon$ arises 
entirely as a result of lensing ($\epsilon=\gamma$) then $\phi_B=0$ and 
$\phi_E$ is equal to the lensing potential $\psi$ (see Straumann 1997).
\\
\indent
Let us introduce the quantities
\begin{equation}
\gamma_E \, = \, 2 \partial \overline{\partial}\phi_E \ , \ \ \ 
\gamma_B \, = \, 2 \partial \overline{\partial}\phi_B .
\end{equation}
It is not difficult to derive, using the complex formalism developed by 
Straumann (1997), the following integral relations for any disk $D$:
\begin{equation}
<\gamma_E >_D \, = \, <\gamma_E>_{\partial D} + <\epsilon_t>_{\partial D} 
\ , \ \ \ 
<\gamma_B >_D \, = \, <\gamma_B>_{\partial D} + <\epsilon_r>_{\partial D} .
\end{equation}
Here $\epsilon_t$, $\epsilon_r$ are the tangential and radial components 
of $\epsilon$,
\begin{equation}
\epsilon_t \, = \, - {\cal R}e \Bigl(\epsilon e^{-2i\varphi} \Bigr) \ , \ \ \ 
\epsilon_r \, = \, - {\cal I}m \Bigl(\epsilon e^{-2i\varphi} \Bigr) , 
\end{equation}
and the averages have to be taken over the disk, respectively over its 
boundary $\partial D$.
\\
In particular, if $\phi_B=0$ we obtain $<\epsilon_r>_{\partial D} = 0$. 
Since for lensing $\gamma_E = \kappa$, we obtain from (13)
\begin{equation}
<\kappa>_{D} \, = \, <\kappa>_{\partial D} + <\gamma_t>_{\partial D} ,
\end{equation}
and 
\begin{equation}
<\gamma_r >_{\partial D} \, = \, 0 .
\end{equation}
The first of these equations is a well-known useful relation between the tangential shear $\gamma_t$ and the projected surface mass density $\kappa$, which 
we shall also use later. Equation (16) provides the test we were looking for (see also Crittenden et al. 2000, and references therein).
\section{Galaxy-Galaxy Lensing}
\label{sec:galaxy}
The main aim of such investigations is the determination of the average mass 
profile of galaxy halos for a population of galaxies to greater distances 
than with conventional methods. I recall that spiral galaxy halos have 
traditionnally been probed via rotation of stars and gas to radii of 
$\sim$ 30 h$_o^{-1}$ kpc. Further away there are no tracers available, but 
since the velocity curves show no decline one expects that the halos are 
considerably more extended. Evidence for this comes also from satellites, 
pairs of galaxies, etc.
\\
\indent
The main problem with lensing by galactic halos is that the signal is at least 
a factor 10 below the noise due to shape variations of the background galaxies.
 However, it is possible to overcome the poor $S/N$ ratio by using a large 
number of lens-source pairs. Such statistical methods can, of course, only 
provide the average of some population (e.g. early 
or late type) of galaxies.
\subsection{Results from Recent Studies}
Let me report on two interrelated observations of galaxy-galaxy lensing which 
supplement each other and led two interesting results.
\\
\indent
Fischer et al. (2000) have measured a galaxy-galaxy lensing signal to a high 
significance from preliminary Sloan Digital Sky Survey (SDSS) data, using a 
sample of $\approx$ 16 million foreground galaxy ($fgg$)/background galaxy 
($bgg$) pairs. These measurements demonstrated the power and potential of 
galaxy-galaxy lensing. The data indicate that halos of typical galaxies 
continue an isothermal profile to a radius of at least 260 h$_o^{-1}$ kpc. 
Unfortunately, there are for the time being only photometric redshift 
distributions available, both for the $fgg$s and the $bgg$s.
\\
\indent
Therefore, the work by Smith et al. (2001) is an important supplement. In this 
measurements of lensing around $fgg$s with redshifts determined by the Las 
Campanas Redshift Survey (LCRS) are reported. Beside the redshift the 
luminosity of the $fgg$s is also known.
\\
\indent
By combining the two sets of data the authors arrive at interesting results. 
First, they obtain average mass profiles in \underline{absolute} units 
(mod $H_o$). Second, the resultant $M/L$, together with the luminosity 
function of LCRS galaxies, gives the galactic contribution to $\Omega_M$. 
The main result of the two papers is that 
\begin{equation}
\Omega_{gal} \, \ge \, 0.2 \ {\rm !}
\end{equation}
\hskip4mm However, another recent study of galaxy-galaxy lensing by Wilson et al. (2000) 
arrives at considerably lower values:
\begin{equation}
\Omega_{halo} \, \simeq \, (0.04 \pm 0.01 ) \ \bigl(r/100 {\rm h}_o^{-1} \, 
{\rm Mpc} \bigr).
\end{equation}
As far as I can see, most of the difference can be traced to different values 
of the Schechter parameter $\phi_\ast$ in the luminosity function. I shall 
say more on this, as well as on other uncertainties, after a more 
detailed discussion of the quoted papers.
\subsection{Detailed Discussion}
Since the lensing signal is so small (less than 1 \%) one has to worry 
about various corrections of the measured galaxy shapes in order to 
determine the true ellipticity. There are, for instance, severe 
difficulties due to atmospheric seeing. Then there are slight anisotropies of 
the telescope (e.g., caused by wind shake).
\\
\indent
If the true ellipticity field is due to lensing it gives the shear field 
$\gamma$, in particular the tangential component centered at many different 
points in the field (for this component some systematics averages out). 
Fischer et al. (2000) give in their Fig. 2 the mean tangential shear around 
$fgg$s measured from images of three filters. A good fit to the data is 
given by a power law:
\begin{equation}
\gamma_T(\theta) \, = \, \gamma_{To} \Bigl( \frac{1^{\prime \prime}}
{\theta} \Bigr)^{\eta} ,
\end{equation}
with parameters given in their Table 2.
\\
\indent
Several tests are applied to verify the reality of the shear detection. I 
only mention the test based on Eq. (16). In practice one just rotates the 
background galaxies by 45$^o$, because under a rotation by an angle $\alpha$ 
the complex shear transforms as $\gamma \rightarrow e^{2i\alpha} \gamma$ 
(as for gravitational waves), and thus $\gamma_t \rightarrow -\gamma_r$ 
for $\alpha = \pi / 4$. It turns out that the signal indeed becomes consistent 
with zero in all three band passes.
\\
\indent
Let me now describe the main steps of the analysis, which is at this stage 
rather primitive. The quality of the data is not yet sufficient to attempt 
a parameter-free reconstruction.
\\
\indent
The surface mass density $\Sigma (\bm{\theta})$ is modelled as 
\begin{equation}
\Sigma \, = \, \Sigma_g \ast \phi , 
\end{equation}
where $\Sigma_g$ is the average of individual galaxies, and $\phi$ takes 
the excess number density of galaxies (due to clustering) into account. For 
$\Sigma_g$ a truncated isothermal mass density, 
\begin{equation}
\rho (r) \, = \, 
\frac{\sigma^2 s^2}{\pi G r^2 (r^2 + s^2)} ,
\end{equation}
is used, where $\sigma$ is the line-of-sight velocity dispersion and $s$ 
is the truncation radius.
\\
\indent
Next, Eq. (15) is used, whereby $\Sigma_{crit}^{-1}$ 
in $\kappa = \Sigma / \Sigma_{crit}$ is replaced by an average, based on the 
photometric redshift data. 
\\
\indent
Finally, a fit to the measured $\gamma_T (\theta)$, Eq. (19), is 
performed. The main result at this stage is that $s>$ 260 h$_o^{-1}$ kpc. 
Since the conversion of shear to mass density relies on photometric 
redshift distributions for the $fgg$s and $bgg$s, we now turn to the work 
by Smith et al. (2001). These authors have measured weak gravitational 
lensing distortions of 450,000 $bgg$s ($20<R<23$) by 790 $fgg$ ($R<18$). The 
latter are field galaxies of known redshift ($0.05<z_f < 0.167$). The 
uncertainties in the $bgg$ redshift distribution turns out not to be 
important ($bgg$s are within reach of current pencil-beam redshift surveys). 
\\
\indent
These data provide the average $<\gamma_t >_{\partial D}$ as a function of 
radius $R$ about galaxies of luminosity $L$ for a population of $bgg$s 
at infinite redshift. If we denote this quantity by 
$\overline{\gamma}_{t,\infty} (R, L, z_f)$, then Eq. (15) gives the relation 
\begin{equation}
\overline{\gamma}_{t, \infty} (R, L, z_f) \, = \, 
\Bigl[ \Sigma_L (\leq R) - \Sigma_L (R) \Bigr] 
\frac{4 \pi G}{c^2} D_f ,
\end{equation}
where $\Sigma_L (R)$ is the mean azimuthally averaged surface mass density 
at radius $R$ about galaxies of luminosity $L$, and 
$\Sigma_L (\leq R)$ denotes the average mass density interior to $R$; 
$D_f$ is the angular diameter distance from the observer to the deflecting 
field galaxy and thus known (mod $H_o$). In principle the square bracket in 
(22) is thus measurable in narrow bins of $L$ and $R$. This is, however, for 
the future. Since the present sample is too small for this, the authors use 
an isothermal profile, for which
\begin{equation}
\Sigma_L (R) \, = \, \Sigma_L ( \leq R)/2 \, = \, 
\frac{v_c^2 (L)}{4 \pi R} .
\end{equation}
The circular velocity $v_c(L)$ is parametrized by a standard power law 
(Tully-Fisher, Faber-Jackson)
\begin{equation}
v_c(L) \, = \, v_\ast (L / L_\ast)^{\beta/2} , 
\end{equation}
where $L_\ast$ is the Schechter parameter in the luminosity function. 
This provides simple scaling laws for $\overline{\gamma}_{T, \infty}$ and 
$M(L)$. The truncation radius is, however, not constraint by the LCRS data, 
and is therefore taken from the SDSS result as $R_{max} >$ 260 h$_o^{-1}$ kpc.
\\
\indent
Fitting the data leads then to the following main results: 
\vskip2mm
(1) The average mass of an $L_\ast$ galaxy inside $R_{max}$ is given by
\begin{equation}
M(L_\ast) (< 260 \, {\rm h}_o^{-1} \, {\rm kpc}) \, = \, 
\left\{ \begin{array}{l}
(3.1\pm 0.8 ) \times 10^{12} \  {\rm h}_o^{-1} \, {\rm M}_\odot 
\ \  \ (\beta = 0.5) , 
\\
(2.7 \pm 0.6)  \times 10^{12} \  {\rm h}_o^{-1} \, {\rm M}_\odot 
\ \ \ (\beta =1.0) .
\end{array} \right.
\end{equation}
(2) The mass/light ratio for the two values of $\beta$ comes out to be 
\begin{equation}
M/L \, = \, 
\left\{ \begin{array}{l}
(360 \pm 90 ) \, {\rm h}_o^{-1} \, {\rm M}_\odot / L_{\odot} \ \ \ 
(\beta = 0.5) , 
\\
(310 \pm 60 ) \, {\rm h}_o^{-1} \, {\rm M}_\odot / L_{\odot} \ \ \
(\beta = 1.0) . 
\end{array} \right.
\end{equation}
\indent
Using also the luminosity function for LCRS galaxies, Smith et al. (2001) find
\begin{equation}
\Omega_{halo} \, \geq \, 
\left\{ \begin{array}{l}
(0.23 \pm 0.06) \ \ \ (\beta = 0.5) , 
\\
(0.16 \pm 0.03 ) \ \ \ (\beta = 1.0) . 
\end{array} \right.
\end{equation}
From this one would conclude that most of the matter in the Universe seems to 
be within 260 h$_o^{-1}$ kpc of normal galaxies.
\\
\indent
Wilson et al. (2000) selected in their data set bright early type galaxies 
($0.1 < z < 0.9)$ and analysed the shear measurements along similar 
lines. They come up with similar $M/L$ ratios for $L_\ast$ galaxies:
\begin{equation}
M/L_B \, = \, (121 \pm 28) \ {\rm h}_o \ 
\Bigl( \frac{r}{100 {\rm h}_o^{-1} \, {\rm Mpc}} \Bigr) .
\end{equation}
When this is translated to the contribution of these halos to the total 
density of the Universe with the 2dF Schechter function fits, the result 
$\Omega_{halo}  = ( 0.04 \pm 0.01) \times (r/100 {\rm h}_o^{-1} \, {\rm Mpc}) 
$ is found.
\\
\indent
$\Omega_{halo}$ depends linearly on the Schechter parameter $\phi_\ast$ and 
this is for the 2dF luminosity function only about half of the one for the 
luminosity function of LCRS galaxies. Numerically, most of the discrepancy with (27) can be traced to this. Such differences of $\phi_\ast$ reflect typical 
uncertainties in the luminosity functions and are presumably hard to overcome. 
\\
\indent
There are a number of other sources of error. For example, the measured 
distortion gets contributions from other mass concentrations along the line 
of sight. It is probably not easy to substract these in order to isolate 
the contribution of the average halo of individual galaxies. The effect of 
galaxy-galaxy correlations is, however, quite small.
\\
\indent
In summary, the studies discussed above have convincingly demonstrated the 
power and potential of galaxy-galaxy lensing. There is agreement that normal 
(early type) galaxies have approximately flat rotation curve halos extending 
out to several hundred h$_o^{-1}$ kpc. It is, however, not yet clear what 
fraction of matter in the Universe is contained within 260 h$_o^{-1}$ 
kpc of normal galaxies. In the foreseeable future we should know more 
about this.
\section*{Acknowledgements}
I wish to thank the ISSI Institute and the local organizers, J. Geiss and 
R. von Steiger, for the opportunity to attend such an interesting workshop.
\section*{References}
{\small
Bartelmann M. et al.: 1998, \newblock {\em Astron. Astrophys.} 
\underline{330}, 1
\\
Bartelmann M., Schneider P.: 2001, \newblock {\em Physics Report} 
\underline{340}, 291
\\
Cheng Y.C., Krauss L.: 2000, \newblock {\em Int. J. Mod. Phys. A} 
\underline{15}, 697, {\em astro-ph/9810393}
\\
Chiba M., Yoshii Y.: 1999, \newblock {\em Astrophys. J} 
\underline{510}, 42
\\
Chiba M., Futamase T.: 1999, \newblock {\em Prog. Theor. Phys. Suppl.} 
\underline{133}, 115
\\
Cooray A.R. 1999, {\em Astron. Astrophys.} \underline{342}, 353
\\
Crittenden R.G. et al. 2000, {\em astro-ph/0012336}
\\
Falco E.E., Kochanek S.K., Munoz J.A. 1998, {\em ApJ} \underline{494}, 47
\\
Fischer P. et al. 1999, {\em Astron. J} \underline{120}, 1198, 
{\em astro-ph/9912119}
\\
Helbig P. et al.: 1999, {\em astro-ph/9904007}
\\
Kaufmann R., Straumann N. 2000, {\em Ann. Phys. (Leipzig)} 
\underline{9}, 384
\\
Kochanek C.S. 1996, {\em Astrophys. J.} \underline{466}, 638
\\
Schneider P., Ehlers J., Falco E.E. 1992, in {\it Gravitational Lenses}, 
{\em Springer, Berlin}
\\
Smith D., Bernstein G., Fischer P., Jarvis M. 2001 {\em ApJ} \underline{551}, 
643, {\em astro-ph/0010071}
\\
Straumann N. 1997, {\em Helv. Phys. Acta} \underline{70}, 894.
\\
Straumann N. 1999, in {\it New Methods for the Determination of Cosmological 
Parameters}, {\em 3$^{eme}$ Cycle de la Physique en Suisse 
Romande} by R. Durrer \& N. Straumann
\\
Wilson G. et al. 2000, {\em astro-ph/0008504}
}
\vskip4mm
\noindent
{\it Adress for Offprints:} N. Straumann, Institute of Theoretical Physics, 
Univesity of Z\"urich, Winterthurerstrasse, 190, CH-8057 Z\"urich-Switzerland, 
norbert@pegasus.physik.unizh.ch
\end{document}